\documentclass[iop]{emulateapj}
\usepackage{color}

\usepackage{epstopdf}
\usepackage{amsmath}

\makeatletter
\newcommand\lsim{\mathrel{\rlap{\lower4pt\hbox{\hskip1pt$\sim$}}
\raise1pt\hbox{$<$}}}
\newcommand\gsim{\mathrel{\rlap{\lower4pt\hbox{\hskip1pt$\sim$}}
\raise1pt\hbox{$>$}}} 
\newcommand{\dm}{\mathrm {dm}}
\newcommand{\br}{\mathrm {b}} 
\newcommand{\com}{\mathrm {com}}

 \newcommand{\tot}{\mathrm{tot}}

\newcommand{\vbc}{\mathrm{vbc}}
\newcommand{\bc}{\mathrm{bc}}

\newcommand{\jcap}{JCAP}

\makeatother
\shorttitle{Baryonic Clumps and Dark halos}
\shortauthors{Naoz \& Narayan }
\makeatother
\begin{document}

\title{Globular Clusters and Dark Satellite Galaxies
  through the Stream Velocity }

\author{Smadar Naoz\altaffilmark{1,2}\altaffilmark{$\dagger$}, \&  Ramesh Narayan\altaffilmark{1}  }
\altaffiltext{1}{ Institute for Theory and Computation, Harvard-Smithsonian Center for Astrophysics, 60 Garden St.; Cambridge, MA, USA 02138}
\altaffiltext{2}{ Department of Physics and Astronomy, University of California, Los Angeles, CA 90095, USA}
\altaffiltext{$\dagger$}{Einstein Fellow}

\email{snaoz@astro.ucla.edu}

\begin{abstract}
The formation of purely baryonic globular clusters with no
gravitationally bound dark matter is still a theoretical challenge. We
show that these objects might form naturally whenever there is a relative
stream velocity between baryons and dark matter. The stream velocity
causes a phase shift between linear modes of baryonic and dark matter
perturbations, which translates to a spatial offset between the two
components when they collapse.  For a $2\sigma$ ($3\sigma$) density
fluctuation, baryonic clumps with masses in the range $10^5-2.5\times
10^6$~M$_\odot$ ($10^5- 4\times10^6$~M$_\odot$) collapse outside the
virial radii of their counterpart dark matter halos. These objects
could survive as long-lived dark matter-free objects and might
conceivably become globular clusters.  In addition, their dark matter
counterparts, which were deprived of gas, might become dark satellite
galaxies.
\end{abstract}

\section{Introduction}\label{intro}

Observations indicate that globular clusters (GCs) contain practically
no gravitationally bound dark matter (DM)
\citep[e.g.,][]{Heggie+96,Bradford+11,Conroy+11,Ibata+13}.  How did
these objects form?  Assuming a baryon-only universe,
\citet{Peebles+68} suggested in early work that GCs formed via
gravitational collapse of non-linear baryonic over-densities shortly
after recombination ($z\sim1000$).  Their model is no longer
viable since we know now that DM dominates the matter content of the
Universe.

\citet{Gunn80} suggested that GCs are formed in strong shocks when gas
is compressed during galaxy mergers. The discovery of many massive
young star clusters in the interacting Antennae system
\citep[e.g.,][]{Whitmore+95,Whitmore+99} supports this idea, and the
scenario has been incorporated in cosmological hierarchical structure
formation models
\citep[e.g.][]{Harris+94,Ashman+92,Kravtsov+05,Muratov+10}.

Another currently popular paradigm is that GCs, like all structure,
initially formed inside DM halos \citep{Peebles84}, but these halos
were later stripped by the tidal field of their host galaxies
\citep[e.g.][]{Bromm+02,Mashchenko+05II,Saitoh+06,Bekki+12}, leaving
the central parts deficient in DM.  However, some GCs are observed
with stellar tidal tails, which is difficult to understand if the
objects have extended DM halos
\citep{Grillmair+95,Moore96,Odenkirchen+03,Mashchenko+05II}.

In the standard model of structure formation, because of
baryon-radiation coupling, baryon over-densities at the time of
recombination were about 5 orders of magnitude smaller than DM
over-densities.  Baryons and DM also had different {\it velocities} at
recombination, with a relative speed $\sim 30~{\rm km\,s^{-1}}$ that
was coherent on comoving scales of a few Mpc \citep{Tes+10a}.  After
recombination, the baryons decoupled from the photons and their
subsequent evolution was dominated by the gravitational potential of
the DM.  They also cooled quickly, and their relative velocity with
respect to the DM, called the ``stream velocity", became supersonic.

The stream velocity has important implications for the first
structures
\citep{Stacy+10,Maio+11,Greif+11,Fialkov+11,Naoz+11a,Naoz+12,OLMc12,BD,Richardson+13,Tanaka+14},
for the redshifted cosmological 21-cm signal
\citep{Dalal+10,Bittner+11,Yoo+11,Visbal+12,McOL12}, and even for
primordial magnetic fields \citep{NN13}.

\citet{Naoz+12} showed that the stream velocity can result in some
halos becoming nearly baryon-free; the gas simply had too much
relative velocity to fall into these DM halos. The question we ask
here is: What happened to the baryons that failed to fall into these
halos?  We show that, in at least some cases, these baryons might have
collapsed to form baryon-only bound objects that are physically
separated from their parent DM halos.  For an interesting range of
masses, the spatial offset is larger than the virial radius of the DM
halo, allowing the baryonic clumps to survive as independent DM-free
objects.  We suggest that these objects may have evolved into GCs. We
also suggest that the corresponding gas-poor DM halos may be present
day dark satellites or ultra-faint galaxies.
 
We begin in \S \ref{sec:Evol} by discussing the evolution of baryonic
over-densities in the presence of a stream velocity. We then calculate
in \S \ref{sec:prob} the likelihood of forming spatially separated
baryon-only objects. We analyze the survival of these objects in \S
\ref{sec:sur} and conclude with a discussion in \S \ref{sec:dis}.
Throughout, we adopt the following cosmological parameters:
$(\Omega_\Lambda,\Omega_m,\Omega_\br,n,\sigma_8,H_0)=
(0.73,0.27,0.044,1.0,0.82,71$~km\,s$^{-1}$Mpc$^{-1}$).

\section{Baryonic over-density and phase shift}\label{sec:Evol}

We solve the coupled differential equations that govern the linear
evolution of the dimensionless density fluctuations of the DM,
$\delta_{\dm}$, and the baryons, $\delta_{\br}$. Both quantities are
complex numbers.  In the baryon frame of reference, the evolution
equations are:
\begin{eqnarray}\label{g_T}
\ddot{\delta}_{\dm} + 2H \dot {\delta}_{\dm}- f_{\dm}\frac{2 i}{a}  {\bf v}_{\bc} \cdot {\bf k} \dot\delta_{\dm}  & = &
 \\
\frac{3}{2}H_0^2\frac{\Omega_{m}}{a^3}
\left(f_{\br} \delta_{\br} + f_{\dm} \delta_{\dm}\right)& +& \left( \frac{  {\bf v}_{\bc} \cdot {\bf k}} {a} \right)^2 \delta_{\dm} \ ,   \nonumber  \\
\ddot{\delta}_{\br}+ 2H \dot {\delta}_{\br}  \ \ \ \ \ \ \ \ \ & = &
  \\
\frac{3}{2}H_0^2\frac{\Omega_{m}}{a^3} \left(f_{\br}
\delta_{\br} + f_{\dm}
\delta_{\dm}\right) &-& \frac{k^2}{a^2}\frac{k_B\bar{T}}{\mu}
\left(\delta_{\br}+\delta_{T}\right) \ , \nonumber \label{eq:db}
 \end{eqnarray}
where $\Omega_m$ is the present day matter density as a fraction of
the critical density, ${\bf k}$ is the comoving wavenumber vector of
the perturbation, ${\bf v}_{\bc}$ is the relative velocity between
baryons and DM in a local patch of the universe, $a$ is the
scale factor of the universe, $H_0$ is the present day value of the
Hubble parameter, $\mu$ is the mean molecular weight of the gas,
$\bar{T}$ is the mean temperature of the baryons, $f_{\br}$
($f_{\dm}$) is the cosmic baryon (DM) fraction, and
$\delta_T$ is the dimensionless fluctuation in the baryon
temperature. Derivatives are with respect to clock time. The above
equations are a compact version of equations (5) in \citet{Tes+10a}.
We have used the fact that $v_{\bc} \propto 1/a$, and have included a
pressure term appropriate to the equation of state of an ideal gas
\citep{NB05}.

The density perturbation amplitudes $\delta_\br$ and $\delta_\dm$ are
complex numbers, with phases given by
\begin{equation}
\phi_{\br,\dm}=\arctan\,[{\rm Im}(\delta_{\br,\dm})/{\rm
    Re}(\delta_{\br,\dm})] \ .
\end{equation} 
The stream velocity introduces a phase shift,
$\Delta\phi=\phi_\br-\phi_\dm$, between the baryons and DM (because of
the third term in the left hand side of Eq.~\ref{g_T}). This phase
shift translates to a physical separation between the DM and baryon
over-densities.  The phase shift was discussed previously by
\citet{Naoz+11a}, but the corresponding spatial shift was not resolved
in their simulations. Here we use analytical linear theory.

\begin{figure}[t!]
\hspace{-0.5cm}
\includegraphics[width=\linewidth]{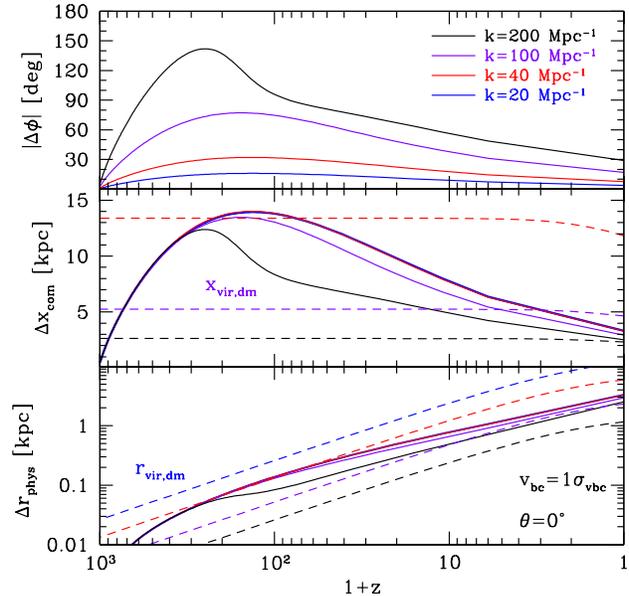} 
\caption{ {\bf Top panel}: Phase shift $\Delta\phi$ between
  fluctuations in the baryons and DM as a function of redshift for
  modes with $k=200,100,40,20$\,Mpc$^{-1}$, assuming a stream velocity
  $v_\bc=1\sigma_\vbc$ and $\theta=0$. {\bf Middle panel}: Comoving
  spatial separation between baryons and DM (solid lines) and comoving
  virial radius of the DM halo (dashed lines). {\bf Bottom panel}:
  Similar to the Middle panel, but shows the {\it physical} spatial
  separation. }\label{fig:phases}
\vspace{0.5cm}
\end{figure}

\begin{figure}[!t]
\hspace{-0.5cm}
\includegraphics[width=\linewidth]{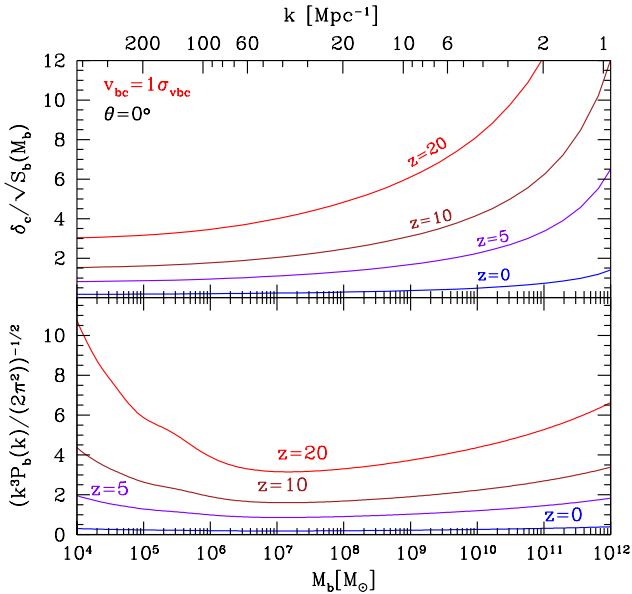} 
\caption{ {\bf Top panel:} The vertical axis measures the rarity of
  fluctuations (``number of $\sigma$'s'') needed to produce a baryonic
  clump of mass $M_\br$ (horizontal axis) at a given redshift 
  (different solid lines); $\delta_c=1.686$ is the critical
  over-density for collapse and $S_\br$ is the variance of $\delta_\br$
  (Eq.~\ref{eq:S}). Results are for a stream velocity
  $v_\bc=1\sigma_\vbc$ and $\theta=0$.  {\bf Bottom panel}:
  Corresponding results for a different measure of fluctuation rarity,
  $1/|\delta_\br|=1/\sqrt{k^3P_\br(k)/(2\pi^2)}$.}\label{fig:sigma}
\vspace{0.3cm}
\end{figure}

Since the phase shift depends on ${\bf v}_{\bc} \cdot {\bf k}$, both
the angle $\theta$ between ${\bf v}_{\bc}$ and ${\bf k}$ and the
magnitude of $v_\bc$ are relevant. For concreteness, we present
results corresponding to $\theta=0$ and $v_\bc=1\sigma_\vbc$, where
$\sigma_\vbc$ is the (scale-independent) rms fluctuation of the stream
velocity on small scales.  In the top panel of Figure
\ref{fig:phases}, we show $\Delta\phi$ as a function of redshift for
four representative wavenumbers, $k = 200,\,100,\,40$ and
$20$~Mpc$^{-1}$, which correspond to baryon masses $M_\br \sim 10^5,
\,10^6, \,10^7, \,10^8$~M$_\odot$, respectively
(Fig.~\ref{fig:sigma}). As seen in Figure \ref{fig:phases}, smaller
scales (larger wave-numbers) develop a larger $\Delta\phi$.  The phase
difference is related to the comoving distance between the baryon and
DM fluctuation peaks, $\Delta x_{\com}$, by
\begin{equation}\label{eq:xvirNom}
\Delta x_{\com}=\left(\Delta \phi/360^\circ \right) 
\left(2\pi/k\right) \ .
\end{equation} 
This comoving separation $\Delta x_\com$, as well as the corresponding
physical separation, $\Delta r_{\rm phys}=\Delta x_\com/(1+z)$, are
shown in the lower two panels of Figure \ref{fig:phases}.

To evaluate how significant the spatial displacement between the baryon
and DM over-densities is, we compare it to the virial radius of the
{\it DM-only} nonlinear object. The virial radius of an
object that collapses at redshift $z$ is approximately
\citep{Bryan+98}:
\begin{equation}\label{eq:r_virBL}
r_{{\rm vir},\dm} \approx 0.784 \left(\frac{M_{\dm}}{10^8 M_\odot
  h^{-1}}\right)^{1/3} \left(\frac{1+z}{10}\right)^{-1} h^{-1} ~ {\rm
  kpc} \ ,
\end{equation}
where, for simplicity, we have suppressed a weak dependence on the
cosmological constant, which causes a slight decrease of the virial
radius at low redshift (this effect is included in the numerical
calculations).  The comoving virial radius is given by $x_{{\rm
    vir},\dm}=r_{{\rm vir},\dm} (1+z)$.

Over a substantial range of $z$, modes with $k=200,\,100$~Mpc$^{-1}$
(or $M_\br \sim10^5, ~10^6M_\odot$) have spatial offsets between their
baryonic and DM linear over-density peaks larger than the DM virial
radius (Fig.~\ref{fig:phases}).  If baryons are able to collapse
at these redshifts, they will form isolated baryonic clumps. In
contrast, modes with smaller values of $k$ (larger $M_\br$) have
separations that lie within the DM virial radius.

\section{ Likelihood of baryon-only clump formation}\label{sec:prob}

When the baryon over-density amplitude $|\delta_\br|$ approaches
unity, the perturbation becomes non-linear and we expect the baryons
to collapse.  To estimate how rare such collapsed objects are we
calculate the variance of $\delta_\br$ as a function of baryon clump
mass $M_\br$,
\begin{eqnarray}\label{eq:S}
S_\br(M_\br,z)&=& \langle |\delta_\br(M_\br,z)|^2 \rangle \\
&=& \int^\infty_0\frac{dk}{2\pi^2}k^2P_\br(k)\bigg[\frac{3j_1(kR)}{kR}\bigg]^2 \ ,\nonumber 
\end{eqnarray}
where $P_\br(k)$ is the spectrum of baryon fluctuations calculated
using Equations (\ref{g_T})--(\ref{eq:db}), $j_1(x)=(\sin x-x \cos
x)/x^2$, and the scale $R$ is the radius of a top-hat window function,
with $M_\br=4\pi \rho_\br R^3/3$.  The value of $S_\br(M_\br,z)$ is
normalized to $\sigma_{8,\tot}(z=0)$ of the total matter. If
$\delta_c$ is the critical linear over-density for collapse, the
rarity of clumps of mass $M_\br$ at redshift is determined by the
number of $\sigma$ fluctuations needed:
$\delta_c/\sqrt{S_\br(M_\br,z)}$ \citep[e.g.][]{BL01,NB07}.  For
simplicity we have set $\delta_c$ to its standard value of $1.686$
\citep[but note that $\delta_c$ may vary with time,
  ][]{NNB,Fialkov+11}.

We estimate the mass $M_\br$ of a baryon-only clump that forms from a
mode with comoving wavenumber $k$ by
\begin{equation}
M_\br=\frac{4\pi}{3}\bar{\rho}_{\br,0}\left(\frac{1}{2}\frac{2\pi}{k}\right)^3=\frac{1}{2}\frac{H_0^2\Omega_\br}{G}\left(\frac{1}{2}\frac{2\pi}{k}\right)^3
\ .
\end{equation}
In the top panel of Figure \ref{fig:sigma} we show the number of
$\sigma$'s by which an over-density of a given baryon mass $M_\br$
must fluctuate in order for it to collapse at a given redshift.
As an example, a $2\sigma$ baryon fluctuation with a mass
$M_\br=10^6$~M$_\odot$ collapses at $z=11.4$. 

The quantity $\delta_c/\sqrt{S_\br(M_\br,z)}$ is a little misleading
since $S_\br$ is computed as an integral over wavenumber $k$. While
the window function cuts off the contribution of all $k$ larger than
about $1/R$, there is no cutoff at low $k$ (large masses). Thus
$S_\br(M_\br)$ has a contribution from fluctuations with masses much
larger than $M_\br$. This contamination is usually not important, but
it is serious when $M_\br$ lies below the Jeans scale so that the
collapse of baryon over-densities is suppressed by gas pressure.  In
this situation, $\delta_c/\sqrt{S_\br(M_\br,z)}$ may indicate collapse
at the $M_\br$ of interest even though what is collapsing is actually
some larger mass that is unaffected by gas pressure.  To illustrate
this point, we show in the bottom panel of Figure \ref{fig:sigma} a
different measure of the rarity of fluctuations,
$1/|\delta_\br|=1/\sqrt{k^3P_\br(k)/(2\pi^2)}$, which focuses on the
local power in baryon density fluctuations at wavenumber $k$. Note the
clear signature of the Jeans cutoff at low masses and high redshifts.

It is important to note that the collapse of baryonic clumps
considered here is very different from collapse in the baryon-only
universe considered by \citet{Peebles+68}. In our case, baryons feel
the gravitational potential of the DM;  their collapse is
driven primarily by the DM. However, because of the stream velocity,
by the time the baryons collapse, they are spatially offset from the
corresponding DM halo.

\begin{figure*}[!t]
\begin{center} 
\includegraphics[width=0.7\linewidth]{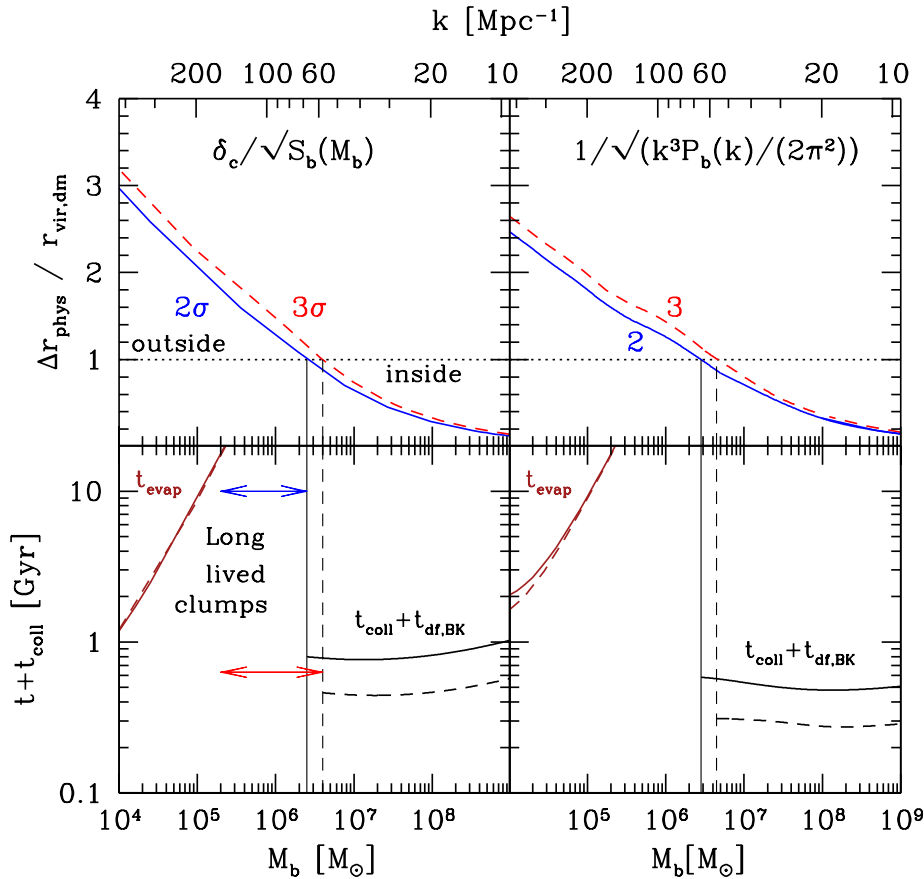} 
\caption{ {\bf Top Left panel:} Spatial separation between baryon and
  DM clumps at the redshift of collapse, normalized by the virial
  radius of the DM halo, for $2\sigma$ (solid line) and $3\sigma$
  (dashed line) fluctuations, plotted against the baryon clump mass
  $M_\br$. Collapse is defined by the condition
  $\delta_c/\sqrt{S_\br(M_\br)}=2,\, 3$, for $2\sigma$ and
  $3\sigma$. Note that clumps with baryon mass larger than
  few$\times10^6M_\odot$ are separated from their DM halos by less than the
  virial radius (they are ``inside'' the DM halo) and vice versa for
  clumps with smaller masses (``outside''). Calculations are for
  $v_\bc=1\sigma_\vbc$, $\theta=0$. {\bf Bottom Left panel:}
  Estimated survival time of baryon clumps against spiral-in and
  merger through dynamical friction (``inside'' the DM halo) and loss
  of stars through evaporation (``outside'' the halo). Baryon clumps
  with masses in the range $\sim10^5-4\times 10^6M_\odot$ are
  potentially long-lived, especially if tidal forces from other nearby
  clumps unbind them from their DM halos.  {\bf Right panels:} Similar
  to Left panels, except that collapse is defined by
  $1/\sqrt{k^3P_\br(k)/(2\pi^2)}=2, \,3$. The results are generally
  similar.}
\label{fig:Delx}
\end{center} 
\end{figure*}

\section{Survival of baryonic clumps}\label{sec:sur}

When a baryonic perturbation collapses, the spatial separation of the
baryon clump from its DM halo is $\Delta r_{\rm phys}= \Delta x_{\rm
  com}/(1+z)$, where $\Delta x_{\rm com}$ is obtained by evaluating
Equation~(\ref{eq:xvirNom}) at the redshift of collapse.  The bottom
panel in Figure~\ref{fig:phases} shows $\Delta r_{\rm phys}$ for some
cases of interest.  The upper panels in Figure~\ref{fig:Delx} show the
same information in a different format. For each $k$, or equivalently
each baryon clump mass $M_\br$, we have identified from
Figure~\ref{fig:sigma} the redshift at which a $2\sigma$
($3\sigma$) fluctuation collapses\footnote{Since GCs account for only
  a small fraction of the baryon content of the universe, they are
  clearly rare, hence we focus on $2\sigma$ and $3\sigma$
  fluctuations, rather than the more common $1\sigma$ fluctuations.},
and computed the ratio of $\Delta r_{\rm phys}$ at this $z$ to the
virial radius $r_{\rm vir,dm}$ (Eq.~\ref{eq:r_virBL}) of the
just-collapsed DM halo. For example, a $2\sigma$ fluctuation with
$M_\br\approx 2.5 \times10^6$~M$_\odot$ collapses with a spatial
offset exactly equal to the virial radius of its DM halo. Clumps with
smaller mass (larger wavenumber) form outside the virial radius, and
vice versa. In the case of a $3\sigma$ fluctuation, the transition
mass is $M_\br\approx 4\times 10^6$~M$_\odot$.

Baryon clumps that form inside the virial radius of the DM host will
spiral down to the center by dynamical friction. To estimate the time
scale we adopt the fitting formula presented in
\citet{Boylan-Kolchin+08}:
\begin{eqnarray}
t_{\rm df,BK}&=&0.22\frac{(M_\dm / M_\br)^{1.3}}
{\ln (1+M_\dm/M_\br)}\frac{\Delta r_{\rm phys}}{r_{\rm vir,\dm}}
e^{1.9\sqrt{1-e_\br^2}} \nonumber \\
&\times&\left(\frac{r_{\rm vir,\dm}^3}{G M_\dm}\right)^{1/2} \ ,
\end{eqnarray}
where $e_\br$ is the eccentricity of the baryon clump's orbit, which
we set to $0.5$.  We add $t_{\rm df,BK}$ to the epoch of collapse
$t_{\rm coll}$ and show the sum in Figure \ref{fig:Delx} (lower
panels).

Baryon clumps that form outside the virial radius are not affected by
dynamical friction. However, these clumps generally have lower masses
and are liable to lose any stars that they form via evaporation.
Following \citet{Gnedin+14} and \citet{Gieles+11}, we estimate the
evaporation timescale as:
\begin{equation}
t_{\rm evap}=17~{\rm Gyr}\frac{M_\br}{2\times 10^5~{\rm M}_\odot} \ .
\end{equation}
Again, we plot the sum $t_{\rm evap}+t_{\rm coll}$.

Figure \ref{fig:Delx} shows that baryon clumps with $M_\br \sim
10^5-4\times 10^6$~M$_\odot$ may be able to survive destruction by
either dynamical friction or evaporation, and may survive as
independent long-lived clumps.  These may be the objects we see today
as GCs. Furthermore, their parent DM halos, which collapsed with a
deficit of baryons, may today be ultra-faint galaxies and dark
satellite galaxies.

While the above proposal seems attractive, could baryon clumps in the
favorable mass range $M_\br \sim 10^5-4\times10^6M_\odot$ fall into
their parent DM halos simply through the gravitational pull of the
latter. The free-fall time from a separation $\Delta r_{\rm phys}$
is very short:
\begin{eqnarray}\label{eq:tff}
t_{\rm ff}&=&\pi\frac{\left(\Delta r_{\rm phys}
  /{2}\right)^{3/2}}{\sqrt{G(M_\br+M_\dm)}}\\ &=& 0.27 {\rm Gyr}
\left(\frac{\Delta r_{\rm phys}}{0.59~{\rm
    kpc}}\right)^{3/2}\left(\frac{\Omega_m}{\Omega_b}\frac{M_\br}{10^6~{\rm
    M}_\odot}\right)^{-1/2} \nonumber \ .
\end{eqnarray}
(Note: a $2\sigma$ fluctuation of $M_\br=10^6~{\rm M}_\odot$ has a
physical separation of $0.59$~kpc from its DM host.)  The actual
timescale is a little longer since the baryon clump will begin with an
outward velocity (Hubble flow).  However, this changes the result by
less than a factor of two.

Newly-formed baryon and DM clumps will certainly free-fall and merge
if they evolve in isolation. More often, however, we expect the two
objects to participate in the hierarchical growth of structure in the
universe. At least some baryon clumps that collapse outside the virial
radius of their DM parent halos will experience strong tidal forces
from neighboring objects and will become gravitationally unbound from
their DM halos. It remains to be seen if a sufficient number can
survive by this mechanism to explain the GCs we observe in the current
universe.\footnote{Note that a previous study by \citet{OLMc12}
  focused on $1\sigma$ fluctuations, which collapse at redshifts well
  below $10$ (according to our analysis), whereas their numerical
  simulation was limited to $z>10$.  Furthermore, because of low
  statistical sampling, they were unable to follow the evolution of
  $2\sigma$ and $3\sigma$ fluctuations. This may explain why they did
  not see any baryon-only clumps such as we predict.}

\section{Discussion}\label{sec:dis}

We have used linear theory to study the growth of baryonic and DM
density fluctuations in the universe in the presence of a stream
velocity $v_\bc$  between the two components. We
focused on the fact that a non-zero stream velocity causes a phase
shift $\Delta\phi$ between the complex amplitudes of the baryonic and
DM density fluctuations, which results in a spatial separation between
the two density peaks. When the perturbations go
non-linear and collapse, the baryon clump forms at a different spatial
location than its DM counterpart. For baryon clump masses less than
about few $\times 10^6M_\odot$, the separation is larger than the virial radius
of the DM halo. Assuming tidal forces from other nearby objects are
able to unbind the baryon clump from its DM halo, the clump could
survive to the present day as a DM-free gravitationally self-bound
object. We suggest that this may be how GCs formed in the
universe. The corresponding baryon-deficient DM halos would similarly
survive as dark satellite galaxies or ultra-faint galaxies, as
suggested previously by \citet{Naoz+12}.

Note that, in this picture, the collapse of a baryon clump is not
driven purely by the self-gravity of the baryons. The primary driving
agent is still the DM perturbation, whose effect on the baryons is
nearly the same as in the standard ($v_\bc=0$) model of structure
formation, so long as the phase shift $\Delta\phi$ is less than about
a radian. For perturbations that satisfy this condition, baryonic
collapse is almost as effective as in the standard model, and gravity
is able to overcome gas pressure; the only difference is that the
baryon and DM clumps form in spatially distinct locations.

In Figure \ref{fig:phases} we showed phase and spatial offsets between
the baryon and DM perturbations for linear modes with different
wavenumbers $k$. As an example, at  $z\approx11.4$, a mode
with $k\approx100$\,Mpc$^{-1}$ (which corresponds to a baryonic mass
$M_\br = 10^6$~M$_\odot$), has a phase shift $\sim 39^\circ$, which
translates to a comoving distance between the baryon and DM density
peaks of 7.3\,kpc. A $2\sigma$ fluctuation with this $k$ collapses at
$z\approx11.4$ (Fig.~\ref{fig:sigma}), and the spatial offset between
the baryonic and DM clumps is about $1.3$ times the virial radius of
the DM halo. For a $3\sigma$ fluctuation, collapse occurs at an
earlier redshift and the separation is even larger.

Using the virial radius of the DM halo as a benchmark spatial
separation to discriminate between baryon clumps that survive and
those that merge through dynamical friction, we obtain a natural upper
cutoff to the mass of baryon-only clumps of a few\,$\times10^6M_\odot$
(Fig.~\ref{fig:Delx}). There is similarly a natural lower cutoff at
around $10^5M_\odot$, which arises from a combination of several
effects: survival against evaporation of stars (Fig.~\ref{fig:Delx}),
Jeans cutoff due to gas pressure, and too large a phase shift between
baryons and DM (which is a serious effect for $k>$ a few hundred
Mpc$^{-1}$). The resulting mass range of long-lived clumps, $M_\br
\sim 10^5-{\rm few}\,\times10^6M_\odot$, agrees well with the masses
of present-day GCs.\footnote{Even if star formation had a relatively
  low efficiency in these pristine gas clumps (say $\sim 10\%$), a
  $4\times 10^6$~M$_\odot$ baryon clump would still have produced
  enough stars to make it an attractive candidate progenitor for
  present-day GCs.  The topic of star formation in our baryon-only
  clumps is beyond the scope of this paper.}  Note that all the
numerical estimates given in this paper are for $v_\bc=1\sigma_\vbc$
and $\theta=0$. The range extends to larger masses for
$v_\bc=2\sigma_\vbc$; $\theta=\pi$ is equivalent to $\theta=0$, but
$\theta=\pi/2$ has no stream velocity effect (within linear theory).

A great deal of work is needed before one can be confident that the
baryon-only clumps discussed here can indeed form naturally and that
they are sufficiently long-lived to be interesting. Within an
analytical approach, one has to  consider
higher-order non-linear techniques.  Alternatively, a numerical
approach is possible, but it will require much better mass resolution
than one usually finds in galaxy formation simulations.  Assuming one
successfully demonstrates that the baryon clumps proposed here form
and are long-lived, one must test  whether or not stars will form inside the
collapsed clumps, and whether a sufficient number of stars will
survive down to $z=0$ with the correct spatial and metallicity
distribution. These are challenging problems.  

Finally we note that, even for baryonic (and DM) masses much larger
than those discussed here (arising from modes with $k\ll10\,{\rm
  Mpc^{-1}}$), there is a significant spatial offset between the
baryons and DM at the time of collapse ($\sim1-3$\,kpc, depending on
the redshift, Fig.~\ref{fig:phases}). These baryon and DM clumps will
quickly merge through dynamical friction. However, in the process, the
baryons will likely cause substantial stirring of the DM fluid in the
inner regions of the halo.  This might well produce a core-like
structure in the DM density distribution, potentially explaining a
number of puzzling observations \citep[for a review,
  see][]{deBlok10}. Also, as Figure~\ref{fig:phases} indicates, the
spatial offset $\Delta r_{\rm phys}$ evolves significantly with the
redshift of formation.  This means that galaxies that formed at higher
redshifts are likely to have smaller dark matter cores, and their
baryonic components will also be more compact, compared to galaxies
that formed more recently
\citep[e.g.,][]{Daddi+05,vanDokkum+08}. These topics are beyond the
scope of this paper.

\section*{Acknowledgments}
We thank Avi Loeb, Rennan Barkana, Steve Furlanetto, Mark Vogelsberger and the  anonymous referee for useful comments.
SN is supported by NASA through an Einstein Post--doctoral Fellowship
awarded by the Chandra X-ray Center,  operated by the
Smithsonian Astrophysical Observatory for NASA, contract
PF2-130096.  RN was supported in part by NSF grant AST1312651.

\bibliographystyle{hapj}


\end{document}